\documentclass[twocolumn,showpacs,preprintnumbers,amsmath,amssymb,prb,floatfix,superscriptaddress]{revtex4}

\usepackage{graphicx,tabularx}
\usepackage{amssymb}
\usepackage{dcolumn}
\usepackage{color}
\usepackage{verbatim}
\usepackage[mathcal]{euscript}
\def\rco{$\mathbf{r}_{\mathrm{CO}}\ $}
\def\rcpt{$\mathbf{r}_{\mathrm{C-surf}}\ $}
\def\ropt{$\mathbf{r}_{\mathrm{O-Pt}}\ $}
\def\aco{$\mathbf{a}_{\mathrm{CO}}\ $}
\def\acpt{$\mathbf{a}_{\mathrm{C-surf}}\ $}
\def\Ef{$\mathrm{E}_{f}\ $}
\def\Ead{$\mathrm{E}_{ad}\ $}

\definecolor{gray0}{gray}{0.0}
\definecolor{gray64}{gray}{0.25}
\definecolor{gray128}{gray}{0.5}
\definecolor{gray192}{gray}{0.75}
\definecolor{gray255}{gray}{1.0}

\begin{document}
\title{CO adsorption on Pt induced Ge nanowires.}
\author{Danny E. P. Vanpoucke}
\altaffiliation[Current Affiliation: ]{Department of Inorganic and Physical Chemistry, Ghent University, Krijgslaan 281 - S3, 9000 Gent, Belgium}
\email[email: ]{dannyvanpoucke@gmail.com}
\homepage{http://users.ugent.be/~devpouck/}
\affiliation{Computational Materials Science, Faculty of Science
and Technology and MESA+ Institute for Nanotechnology, University
of Twente, P.O. Box 217, 7500 AE Enschede, The Netherlands}
\author{Geert Brocks}
\affiliation{Computational Materials Science, Faculty of Science and Technology and MESA+ Institute for Nanotechnology, University of Twente, P.O. Box 217, 7500 AE Enschede, The Netherlands}

\date{\today}
\begin{abstract}
Using density functional theory, we investigate the possible adsorption sites of CO molecules on the recently discovered Pt induced
Ge nanowires on Ge(001). Calculated scanning tunneling microscope (STM)
images are compared to experimental STM images to identify the
experimentally observed adsorption sites. The CO molecules are found
to adsorb preferably onto the Pt atoms between the Ge nanowire dimer segments.
This adsorption site places the CO molecule in between two nanowire dimers, pushing
them outward, blocking the nearest equivalent adsorption sites. This
explains the observed long-range repulsive interaction between CO
molecules on these Pt induced nanowires.
\end{abstract}

\maketitle
\section{Introduction}
In the last several decades, CO adsorption on Pt surfaces has been
studied extensively both experimentally and theoretically. This large
interest is partly due to the deceiving simplicity of the system and
its industrial importance in catalytic processes, such as CO oxidation
and Fischer-Tropsch synthesis.\cite{Eichler:ss02,Ertl:ChemRev95}\\
\indent However, a simple system such as CO adsorbed on the
Pt(111)-surface, has and still does cause quite some controversy. Three
decades ago, adsorption site preference and measured adsorption energies
were the subject of discussion among experimental researchers. These
problems have meanwhile been resolved and experimental results
have converged to a coherent and detailed picture of this
system.\cite{Ertl:ss77,Froitzheim:ap77,Steininger:ss82,Ogletree:ss86,Yeo:jpchem97}
On the theorists side however, a discussion has emerged during the last
decade regarding the unexpected failure of prevalent density functional theory (DFT) approximations
to properly predict the CO/Pt(111) site preference. From experiment it
is found that the ontop site is most stable in the low density regime,
while local density approximation (LDA) and generalized gradient approximation (GGA) calculations show a preference for the threefold
coordinated hollow adsorption site. The cause of this CO/Pt(111) puzzle seems
to originate from the tendency of LDA and GGA to favor higher coordination
and the flatness of the potential surface describing adsorption of CO on the Pt(111) surface.\cite{Feibelman:jpcb01} This has lead to a search for better or alternative functionals in recent years.\cite{Beurden:prb02,Alaei:PRB08}\\
\indent Although the incorrect site prediction is a problem for DFT,
this does not mean that the obtained geometries and derived physical
properties are incorrect.\cite{Dabo:jacs07} Even more, the calculated
STM images derived from the geometries show excellent qualitative
agreement with the experiment.\cite{Bocquet:ss96,Pedersen:cpl99}\\
\indent With the recent discovery of Pt induced nanowire (NW) arrays on Ge(001), a new Pt based adsorption surface becomes available.\cite{Gurlu:apl03} Decoration of these NWs with CO-molecules opens the way to the
formation of one-dimensional ($1$D) molecular chains. Although the adsorption
of single CO-molecules on these Pt induced NWs has been observed
experimentally, true molecular chains remain to be
observed.\cite{Oncel:ss06,Kockmann:prb08}\\
\indent Room temperature (RT) STM experiments, by \"Oncel \textit{et al.}\cite{Oncel:ss06}, showed the CO molecules to be very mobile along the NWs. Later, Kockmann \textit{et al.}\cite{Kockmann:prb08} performed experiments at $70$ K to suppress this mobility, and observed a long range
repulsive interaction between pairs of CO molecules on the same NW.
In those experiments the NWs were considered to
be composed of Pt dimers in the troughs of a modified Ge(001) surface,
called $\beta$-terrace,\cite{Gurlu:apl03} allowing for a straight
forward interpretation of the observed STM images. The CO molecules
were suggested to be adsorbed on the bridge positions of
the NW dimers, comparable to the adsorption of CO on the Pt(001) surface.\cite{Oncel:ss06,Kockmann:prb08}\\
\indent Calculations on the interaction of CO with a free standing Pt
monatomic wire suggest a similar behavior.\cite{Tosatti:prb08} However,
in recent theoretical studies we showed the NWs to be modeled by Pt induced Ge NWs.\cite{vanpoucke:prb2008R,Vanpoucke:prb09NW} In this model the NWs consist of Ge dimers placed in the Pt lined troughs of a Pt modified Ge(001) reconstructed surface. Since the sticking probability and affinity for CO on Ge is known to be low,\cite{Fukutani:jesrp98} while being high for Pt, it would be surprising if CO molecules would adsorb on the Ge NW itself. This might lead to the suggestion that the theoretical models, proposed in Ref.~\onlinecite{Vanpoucke:prb09NW}, are in
disagreement with the experiment. How can this theoretical model be reconciled
with the experimental observations?\\
\indent In this paper we study the adsorption of CO on Pt induced
NWs, starting from the theoretical models we proposed
previously in Ref.~\onlinecite{vanpoucke:prb2008R} and \onlinecite{Vanpoucke:prb09NW}. Using \textit{ab initio} DFT calculations, formation and adsorption energies are calculated. Theoretical STM images, generated using the Tersoff-Hamann method, are compared to experimental STM images to identify the adsorption sites and geometries observed in
experiment.\\
\indent This paper is structured as follows: In
Sec.~\ref{sc:theormeth} the used theoretical methods are described. In Sec.~\ref{sc:results} we present our results, which will be discussed
more in depth in Sec.~\ref{sc:discusion}. Finally, in
Sec.~\ref{sc:conclusion} the conclusions are given.
\section{Theoretical method}\label{sc:theormeth}
The calculations are performed within the DFT framework using the
projector augmented waves method and the Ceperley-Alder LDA functional, as implemented in the VASP program.
\cite{Blochl:prb94,Kresse:prb99,Kresse:prb93,Kresse:prb96} A $400$ eV kinetic
energy cutoff is applied for the plane wave basis set. CO molecules are placed on the models of both types of Pt induced
NWs on Ge(001) we presented in Ref.~\onlinecite{Vanpoucke:prb09NW}.
The surface/NW system is modeled by periodically repeated slabs of $12$
layers of Ge atoms with NW reconstructions on both surfaces. A vacuum
region of $\sim15.5$ \AA\ is used to separate the periodic images of the slab
along the $z$ axis. Due to the computational cost and the small size of the CO
molecule a $(2\times4)$ surface cell for the solitary wire geometry, and a $(4\times4)$ surface cell for the array wire geometry is used. The Brillouin zone of the $(2\times4)$ ($(4\times4)$) surface unit cell is sampled using a
$8\times4$ ($4\times4$) Monkhorst-Pack special $k$-point
mesh.\cite{Monkhorst:prb76} To optimize the geometry of the
surface/adsorbate system the conjugate gradient method is used while the
positions of the Ge atoms in the center two layers are kept fixed as to
represent the bulk of the system.\\
\indent STM images are calculated using the Tersoff-Hamann method in its most
basic form, with the STM tip approximated as a
point-source.\cite{Tersoff:prb85} The integrated local density of states (LDOS) is calculated as
$\overline{\rho}(\mathbf{r},\varepsilon)
\propto \int_{\varepsilon}^{\varepsilon_{\mathrm{F}}}
\rho(\mathbf{r},\varepsilon^{\prime})\mathrm{d}\varepsilon^{\prime}$,
with $\varepsilon_{\mathrm{F}}$ the fermi energy.  Because the
tunneling current is proportional to the integrated LDOS in the Tersoff-Hamann model, an STM-tip following a surface of constant
current can be simulated through plotting a surface of constant (theoretical)
LDOS: $\overline{\rho}(x,y,z,\varepsilon)=\mathrm{C}$, with C a
constant. For each C this construction returns a height $z$ as a
function of the position $(x,y)$. This heightmap is then mapped
linearly onto a gray scale. The constant C is chosen such that the isosurface
has a height $z$ between $2$ and $3$ \AA\ above the highest atom of the
surface.
\section{Results}\label{sc:results}
As is shown in literature, a small difference exists between solitary
NWs (NW$1$) and NWs in arrays (NW$2$).\cite{fn:def_solNW} In experiment this difference presents itself as the appearance of a $(4\times1)$ periodicity at lower temperatures, which was traced back, in previous calculations,
to the presence of an extra Pt atom bound to every pair of NW
dimers.\cite{Houselt:ss08,Vanpoucke:prb09NW}
Since this extra Pt atom introduces new possible adsorption sites and
geometries, CO adsorption on both NW geometries is studied.\\
\indent Because some of the initial adsorption geometries relaxed into
the same final structure, and because in some cases the geometry was
modified extensively during relaxation, the adsorption sites presented
in this manuscript are those found after relaxation.\\
\indent The NW geometry is a metastable configuration and the
adsorption of CO sometimes introduces large deformations of the
surface. Therefore we define both a formation and adsorption energy in
these systems. The formation energy \Ef indicates the energy gain/loss
of the entire system due to the CO adsorption and the subsequent
changes in the surface structure. It is defined (per
$(4\times2)$-surface unit cell) as:
\begin{equation}\label{eq:1_CO_Ef}
\mathrm{E}_{f} = (\mathrm{E}_{\mathrm{NW+CO}} - \mathrm{E}_{\mathrm{prist NW}}
-
\mathrm{N}_{\mathrm{CO}}\mathrm{E}_{\mathrm{CO}})/2,
\end{equation}
with $\mathrm{E}_{\mathrm{NW+CO}}$ the total energy of the
adsorbate-surface system, $\mathrm{E}_{\mathrm{prist NW}}$ the total
energy of a pristine slab $+$ NW system and $\mathrm{E}_{\mathrm{CO}}$ the
total energy of a free CO molecule. $\mathrm{N}_{\mathrm{CO}}$ is
the number of CO molecules per surface unit in the system and the
division by two is because CO is adsorbed at both faces of the slab. A
negative value of the formation energy \Ef indicates an increase in
stability of the system. The adsorption energy \Ead refers to the
binding energy of the CO molecule to the surface. Here any contribution
due to surface deformation is excluded. It is defined (per CO molecule)
as:
\begin{equation}\label{eq:2_CO_Ead}
\mathrm{E}_{ad} = (\mathrm{E}_{\mathrm{NW+CO}} -
\mathrm{E}_{\mathrm{NW+sd}} -
\mathrm{N}_{\mathrm{CO}}\mathrm{E}_{\mathrm{CO}})/(2\mathrm{N}_{\mathrm{CO}}),
\end{equation}
with $\mathrm{E}_{\mathrm{NW+sd}}$ the total energy of the surface with
the adsorption induced deformations but without adsorbed CO molecule.

\subsection{CO on solitary NWs}\label{ssc:res_NW1}
\begin{figure}[!tb]
\begin{center}
 \includegraphics[width=8cm,keepaspectratio]{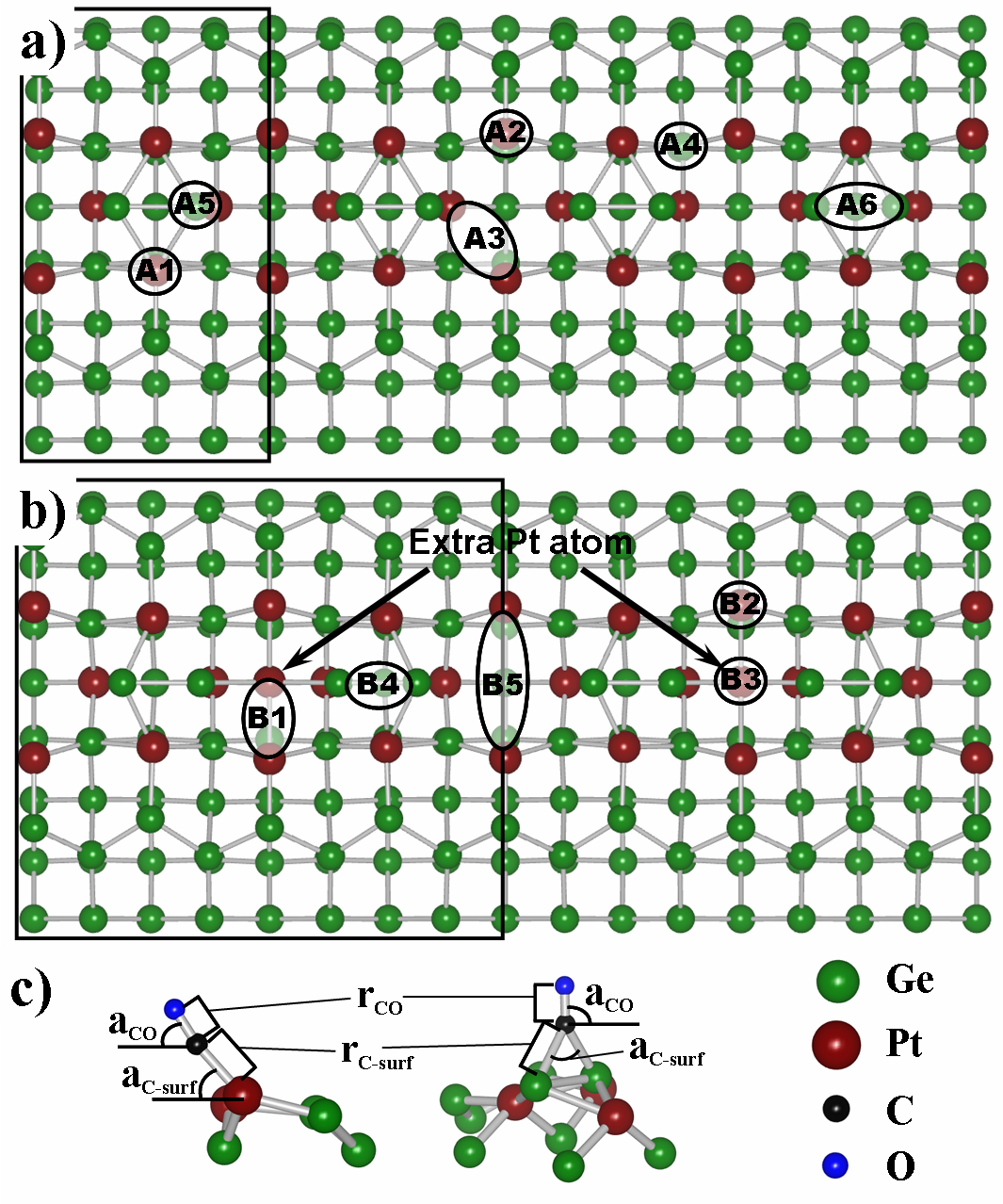}\\
\end{center}
\caption{(Color online) Adsorption sites of CO on a solitary NW geometry (a) and an
array NW geometry (b). Green (red) balls indicate the positions of the
germanium (platinum) atoms. Black rectangles indicate surface unit
cells.}\label{fig:1NWadsorbsites}
\end{figure}

\begin{figure}[!tb]
\begin{center}
\includegraphics[width=8cm,keepaspectratio]{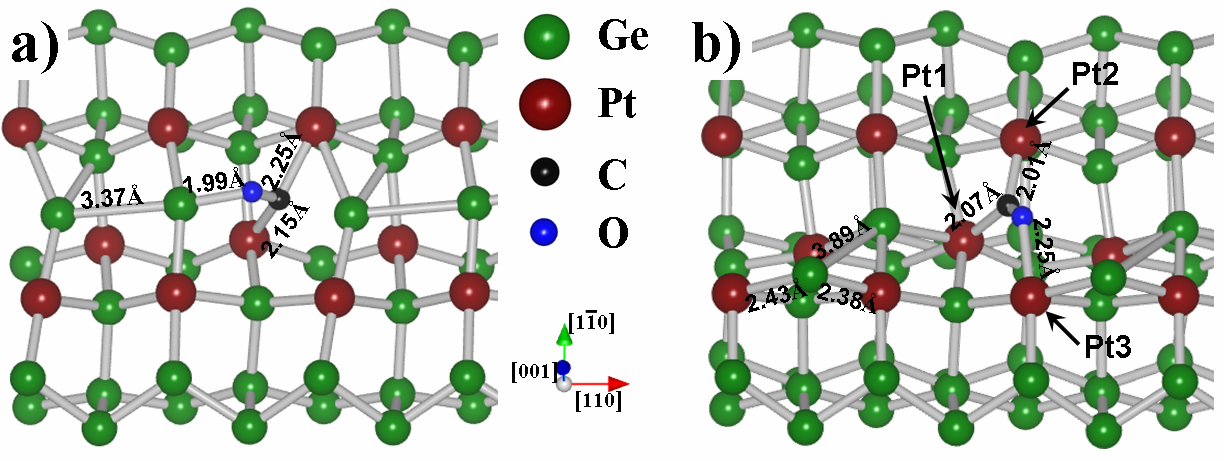}\\
\end{center}
  \caption{(Color online) Ball-and-stick representations of the relaxed A$3$a (a) and A$3$b
  (b) adsorption structures.}\label{fig:2A3akat3geom}
\end{figure}

Solitary NWs consist of Ge dimers located in the Pt lined troughs of a
Pt modified Ge(001) surface. We will refer to this structure as NW$1$. Figure~\ref{fig:1NWadsorbsites}a shows the
adsorption sites studied for this NW1 surface reconstruction.
The adsorption and formation energies are given in
Table~\ref{table:1energeom_NW1}, as are some geometrical parameters, defined in Fig.~\ref{fig:1NWadsorbsites}c, for the CO molecule on the surface.
\begin{table}[!t] \center{\textbf{Adsorption energies and geometrical
parameters\newline for CO on NW$1$. }}
\begin{ruledtabular}
\begin{tabular}{l|rrcrcrc}
   & \multicolumn{1}{c}{\Ef} & \multicolumn{1}{c}{\Ead} &
   \multicolumn{1}{c}{coord.} & \multicolumn{1}{c}{\rco} &
\multicolumn{1}{c}{\aco} & \multicolumn{1}{c}{\rcpt} &
\multicolumn{1}{c}{\acpt} \\
   & \multicolumn{1}{c}{(eV)} & \multicolumn{1}{c}{(eV)} & &
   \multicolumn{1}{c}{(\AA)}&
\multicolumn{1}{c}{($^{\mathrm{\circ}}$)} & \multicolumn{1}{c}{(\AA)} &
\multicolumn{1}{c}{($^{\mathrm{\circ}}$)} \\
  \hline
  NW$1$ A$1$ & $-1.300$  & $-1.747$ & t & $1.152$ & $88$ & $1.906$ & $86$  \\ 
  NW$1$ A$2$ & $-1.721$  & $-1.990$ & t & $1.154$ & $57$ & $1.920$ & $48$  \\ 
  NW$1$ A$3$a& $-1.156$ & $-0.993$ & b & $1.291$ & $56$ & $2.246$ & $117$ \\
  & & & & & & ($2.150$)& \\ 
  NW$1$ A$3$b& $-1.485$ & $-2.069$ & b & $1.214$ & $47$ & $2.012$ & $101$  \\
  & & & & & & ($2.066$)& \\
  NW$1$ A$4$ & $-2.859$  & $-1.890$ & t & $1.152$ & $63$ & $1.907$ & $62$  \\ 
  \hline 
  NW$1$ A$5$ & $-1.421$  & $-0.439$ & t & $1.151$ & $70$ & $2.015$ & $70$  \\ 
  NW$1$ A$6$ & $-0.349$  & $-0.557$ & b & $1.162$ & $88$ & $2.192$ & $75$  \\ 
\end{tabular}
\end{ruledtabular}
\caption{Formation and adsorption energies for CO adsorbed on the NW1-surface. Adsorption sites are shown in Fig.~\ref{fig:1NWadsorbsites}a. t (b)
refers to top (bridge) adsorption.
\rco and \rcpt are the C-O and C-Pt(Ge) bond lengths. For the A3
adsorption site the value between brackets is the C-Pt bond length
between C and the Pt atom in the bottom of the trough. \aco and \acpt
are the bond angles with regard to the surface plane. In case of bridge
adsorption \acpt is the angle between the two C-surface bonds. For the
adsorption sites A1 to A4, the C atom is bound to surface Pt atoms, for the sites A5 and A6 the C atom is bound to Ge NW atoms.}\label{table:1energeom_NW1}
\end{table}
The dimer length \rco of a free CO molecule was calculated to be
$1.1330$ \AA, in good agreement with the experimental
value.\cite{Gilliam:pr50} Table~\ref{table:1energeom_NW1} shows the CO
bond lengths are only slightly stretched in most cases: about $1.59-1.85$\%. The exception
being CO adsorbed in the A$3$ configurations where the stretching is
$13.95$\% and $7.15$\%. The difference between these last two configurations
and the other configurations is the extra bond of the O atom with one of the
surface atoms. In case of the A3a configuration the O atom has an extra
bond with a Ge dimer atom [\textit{cf.}\ Fig.~\ref{fig:2A3akat3geom}a], while in the A3b configuration it is bound to a Pt atom of the top layer at the
opposite side of the trough [\textit{cf.}\ Fig.~\ref{fig:2A3akat3geom}b]. Another interesting geometrical feature is that in most cases the adsorbed CO molecule is tilted with regard to the surface, unlike the behavior of CO molecules on clean Pt surfaces. In case of CO adsorbed on top of Pt atoms, the C-Pt bond length is just slightly longer than what is found for CO on Pt(111) in an ontop configuration.\cite{Hirschl:prb02,Eichler:ss02}\\
\indent The adsorption energies given in Table~\ref{table:1energeom_NW1} show a very clear preference
for CO adsorption on Pt (sites A$1$--A$4$). The values of \Ead might indicate that CO also binds weakly to the Ge NW atoms (sites A$5$ and A$6$), contrary to the experimental knowledge that CO does not bind to Ge. However, one needs to bare in mind that LDA tends to overbind, which in this case results in the small adsorption energies. Furthermore, the binding energies of CO on Ge are
\Ead$\cong 0.5$ eV, while \Ead$\cong2.0$ eV for CO adsorbed on Pt atoms,
making the latter much more preferable.\\
\indent The A$3$-structures are a bit peculiar, and, as we will show later, the A$3$a-structure results in calculated STM images that show extremely good agreement with the STM images obtained by \"Oncel \textit{et al.}\cite{Oncel:ss06} Both A$3$-structures have a bridge like
adsorption geometry [\textit{cf.}\ Fig.~\ref{fig:2A3akat3geom}] but
their adsorption energies differ more than $1$ eV, making them the best
and worst cases for CO adsorbed on the Pt atoms in this system.
The  most important contribution  to the adsorption energy in this system comes from the CO bond-stretching. To examine the effects of this bond stretching, one can use a modified version of Eq.~\eqref{eq:2_CO_Ead}, replacing the formation energy of a relaxed gas molecule by the formation energy of a CO gas molecule with the same bond length as the adsorbed molecule. The resulting energy could then be considered the energy needed to desorb the molecule from the surface in a two step fashion. First, the bond(s) to the surface atoms are broken while the CO bond length remains unchanged. Then the molecule moves away from the surface relaxing to its equilibrium bond length. Since most adsorption geometries show only slightly stretched CO molecules, this energy only differs a little ($\sim50$ meV) from \Ead. However, for the strongly stretched A$3$-geometries a large difference with regard to \Ead is seen, resulting in energies of $-2.010$ eV and $-2.384$ eV for the A$3$a- and A$3$b-structures, respectively. This means that a CO molecule at an A$3$-site becomes the hardest to desorb from the surface using the path described above.\\
\indent The somewhat peculiar adsorption geometry in case of the A$3$b-site can be understood when it is compared to the adsorption of CO on a
monatomic Pt wire. In their theoretical study, Sclauzero \textit{et
al.}\cite{Tosatti:prb08} found that for an intermediate
inter-platinum-distance of $3.7$--$5.3$ \AA\ a configuration with the CO molecule bridging the gap between two Pt atoms is the most stable configuration, while for an inter-platinum-distance around $2.50$ \AA\ the bridge configuration, with the C atom bound to two Pt atoms, was found to be most favorable. In the A$3$b-situation three Pt atoms
are involved. Two (Pt$1$ and Pt$2$) are bound to the C atom and one
(Pt$3$) is bound to the O atom. The distance between Pt$1$ and Pt$2$ is
$3.145$ \AA, while the distance between Pt$1$ and Pt$3$, and Pt$2$ and
Pt$3$ is $3.958$ and $5.008$ \AA, respectively. Both the latter values are
nicely in the intermediate range for the bridging configuration, while the
former value lies in the range where Sclauzero \textit{et al.} found a simple bridge configuration to be most stable.\\
\begin{table}[!t] \center{\textbf{CO on the $4\times$ NW1 surface cell.}}
\begin{ruledtabular}
\begin{tabular}{l|ccccccc}
   & A$1$ & A$2$ & A$3$b & A$4$ & A$5$ & A$6$ \\
  \hline
  \Ef (eV)  & $-3.60$  & $-3.58$ & $-5.56$ & $-2.72$ & $-0.78$ & $-1.35$ \\
  \Ead (eV) & $-1.94$  & $-2.11$ & $-2.00$ & $-1.94$ & $-0.67$ & $-0.39$ \\
\end{tabular}
\end{ruledtabular}
\caption{Adsorption and formation energies for the adsorption of one CO
molecule per four NW dimers.}\label{table:2ener_NW1mega}
\end{table}
\indent From Eqs.~\eqref{eq:1_CO_Ef} and \eqref{eq:2_CO_Ead} follows that the energy due to the surface deformation induced by the adsorption of a CO molecule is given by: $\mathrm{E}_{sd}=\mathrm{E}_{f}-\mathrm{E}_{ad}$. Positive values of $\mathrm{E}_{sd}$ indicate a destabilization due to the adsorbed CO molecule (A$1$, A$2$, A$3$b, and A$6$), while negative values indicate a stabilization (A$3$a, A$4$, and A$5$). In Ref.~\onlinecite{Vanpoucke:prb09NW} the NW$1$-geometry was shown to be a metastable configuration. It was also shown that the adsorption configuration with the Ge NW dimer bound to four surface dimers (site C in Fig.~$11$a in Ref.~\onlinecite{Vanpoucke:prb09NW}) has a more favorable formation energy than the NW$1$ structure (\textit{cf.}\ Table~III in Ref.~\onlinecite{Vanpoucke:prb09NW}, compare $\gamma_{as}^{\star}$ Ge NW A (NW$1$) to $\gamma_{as}^{\star}$ Ge NW C).\\
\indent This explains the increased surface stability seen for CO adsorption sites A$3$a, A$4$ and A$5$. From the large differences between the adsorption and formation
energies, it is expected that adsorption of CO on the NW$1$-system has a
large influence on the geometry of the wire, as can be seen for example in Fig.~\ref{fig:2A3akat3geom}. Because the
range of the (destructive) influence of a CO molecule on the wire is larger
than the size of the used surface cell, some calculations are carried out
using a surface cell $4\times$ the unit cell length. Only a single CO
molecule is absorbed on it, effectively reducing the CO density by a
factor of four. Since the relaxation of such a huge cell is
computational very demanding, lower accuracy relaxation parameters are
used. \Ef and \Ead are given in Table~\ref{table:2ener_NW1mega} showing even more clearly the effect of the CO molecule on the NW$1$ geometry\\
\indent The large differences between \Ef and \Ead again indicate large
modifications of the NW$1$-structure. Comparing the relaxed
geometries, we find that for larger differences between \Ef and
\Ead, more NW dimers are displaced or destroyed. In case of the
A$3$b-structure even all four NW dimers are somehow modified. This
behavior might be the basis of the experimentally observed long-ranged
repulsive interaction between CO molecules.\cite{Kockmann:prb08} We will look
into this aspect of CO adsorption in more detail in Sec.~\ref{sc:discusion}.\\
\begin{table}[!t] \center{\textbf{Adsorption energies and geometrical
parameters\newline for CO on NW$2$.}}
\begin{ruledtabular}
\begin{tabular}{l|rrcrcrc}
   & \multicolumn{1}{c}{\Ef} & \multicolumn{1}{c}{\Ead} &
   \multicolumn{1}{c}{coord.} & \multicolumn{1}{c}{\rco} &
\multicolumn{1}{c}{\aco} & \multicolumn{1}{c}{\rcpt} &
\multicolumn{1}{c}{\acpt} \\
   & \multicolumn{1}{c}{(eV)} & \multicolumn{1}{c}{(eV)} & &
   \multicolumn{1}{c}{(\AA)}&
\multicolumn{1}{c}{($^{\mathrm{\circ}}$)} & \multicolumn{1}{c}{(\AA)} &
\multicolumn{1}{c}{($^{\mathrm{\circ}}$)} \\
  \hline
  NW$2$ B$1$ & $-1.424$  & $-2.336$  & b & $1.171$ & $76$ & $2.086$ & $82$ \\ 
  &  &  &  &  &  & ($2.004$) &  \\
  NW$2$ B$2$ & $-0.761$  & $-1.911$  & t & $1.154$ & $77$ & $1.911$ & $74$ \\ 
  NW$2$ B$3$ & $-1.033$  & $-2.090$  & t & $1.155$ & $90$ & $1.868$ & $90$ \\ 
  NW$2$ B$5$ & $-0.784$  & $-2.197$  & t & $1.170$ & $20$ & $1.883$ & $28$ \\ 
  \hline 
  NW$2$ B$4$ & $-0.319$  & $-0.916$  & b & $1.163$ & $62$ & $2.003$ & $72$ \\ 
  &  & & & &  & ($2.411$) & \\
\end{tabular}
\end{ruledtabular}
\caption{Formation and adsorption energies for CO adsorbed on the NW2
surface. Adsorption sites are shown in Fig.~\ref{fig:1NWadsorbsites}b.
t (b) refers to top (bridge) adsorption. In case of the B5 adsorption
site, the C atom is bound to only one Pt atom, however, the O is bound
to the Pt atom at the opposite side of the trough resulting in the
entire CO molecule forming a bridge. \rco and \rcpt are the C-O and
C-Pt(Ge) bond lengths. \aco and \acpt are the bond angles with regard
to the surface plane. For the B1 adsorption site the value between
brackets is the C-Pt bond length between C and the Pt atom at the
bottom of the trough. For the B4 adsorption site the value between
brackets is the C-Ge bond length to the Ge atom bound to the extra Pt
atom in the trough. In case of bridge adsorption \acpt is the angle
between the two C-surface bonds. Only for the B4 site C is bound to
Ge NW atoms, in all other cases C is bound to a Pt atom in the surface.
}\label{table:3energeom_NW2}
\end{table}

\subsection{CO on NW-arrays}\label{ssc:res_NW2}
In Ref.~\onlinecite{Vanpoucke:prb09NW} it was shown that the NWs in NW-arrays
only differ very little from solitary NWs: a single extra Pt atom in the NW
trough added to two NW$1$ unit cells, we will refer to this structure as NW$2$. The extra Pt atom (indicated in
Fig.~\ref{fig:1NWadsorbsites}b) binds to two NW dimers, inducing the
observed $4\times1$ periodicity. This extra bond also stabilizes the
NW.\cite{Vanpoucke:prb09NW,Vanpoucke:mrs09eprocNW} Through its construction, the unit cell of
a NW$2$ surface is twice the size of that of the NW$1$ surface and
since we only adsorb $1$ CO molecule per surface, this effectively
halves the CO-coverage compared to that on the NW$1$ system.\\
\indent Figure~\ref{fig:2A3akat3geom}b shows the adsorption sites of CO on the NW$2$
$4\times4$ surface cell after relaxation of the system.\cite{fn:fig2surf}
Only the adsorption sites different from those already
present in the NW$1$-geometry, \textit{i.e.}\ those where the extra Pt atom is involved directly of indirectly, are investigated. Our observations for the adsorption sites on the NW$1$-surface are assumed also to be
valid for the equivalent sites on the NW$2$-surface.\\
\indent The adsorption energies per CO molecule and the formation
energies per $4\times2$ surface cell are shown in
Table~\ref{table:3energeom_NW2}. Just as for the the CO molecules adsorbed on
Ge in the NW$1$ case, CO molecules adsorbed on Ge in the NW$2$ case (B$4$)
have a much lower binding energy than CO molecules bound to Pt atoms,
indicating the strong preference of CO toward Pt. The adsorption
energies for CO molecules on Pt atoms are in the same range as for the
NW$1$-system, albeit slightly higher on average. This
might be due to the lower CO coverage in the NW$2$-systems, reducing the direct and indirect interaction between CO molecules. The A$2$
and B$2$ adsorption sites, on the NW$1$- and NW$2$-geometries,
respectively, differ only by the presence of the extra Pt atom,
resulting in comparable adsorption energies. However, the presence of
the extra Pt atom in case of the NW$2$-system prevents the CO molecule
of bending far toward the surface, increasing the angle \acpt from
$48^{\circ}$ to $74^{\circ}$.\\
\indent In case of the B$5$-site, where no extra
Pt atom is present in the trough, the CO molecule bends very much
toward the surface and actually bridges the through connecting two Pt
atoms at opposing sides of the trough.\\
\indent The distance between these Pt atoms is $4.826$ \AA, placing them in the regime where Sclauzero \textit{et al.}\cite{Tosatti:prb08} found a ``\textit{tilted bridge}'' configuration to be the energetically preferred configuration. In their study of CO adsorption on a freestanding monatomic Pt wire, Sclauzero \textit{et al.} identified $3$ regimes for the CO adsorption geometry. i) For an unstretched (with Pt-Pt bond length $\mathrm{d}_{\mathrm{Pt-Pt}}=2.34$ \AA) freestanding Pt wire, the bridge configuration was found to be favored with respect to the ontop configuration by about $1$ eV. The energy of the bridge configuration was shown to have a minimum for a value for $\mathrm{d}_{\mathrm{Pt-Pt}}=2.50$ \AA\ just slightly above the equilibrium Pt bond length of the chain. We will refer to this as the unstretched regime. ii) In case of a hyperstretched configuration an energy minimum was found at $\mathrm{d}_{\mathrm{Pt-Pt}}=5.05$ \AA. In this case the substitutional geometry was found to be more favorable than the bridge configuration. In the substitutional configuration the CO molecule is aligned parallel with the Pt wire, and the C and O atom are bound each to a half of the Pt wire. iii) However, Sclauzero and his collaborators found the energy minimum of the substitutional geometry to be still slightly higher than the tilted bridge configuration. In this tilted bridge configuration the CO bond lies in a plane trough the wire, but it is not aligned parallel or orthogonal to the wire. Although no energy minimum for this configuration was found, Sclauzero \textit{et al.} found it to be preferred over the bridge and substitutional configuration for intermediate stretching lengths of the Pt-Pt distance (about $3.8$ \AA\ $\leq \mathrm{d}_{\mathrm{Pt-Pt}} \leq 5.1$ \AA).\\
\indent Although the monatomic wire studied by Sclauzero \textit{et al.} are quite different from our current system, the inter-platinum-distances are comparable. Even more, in case of the NW$2$-geometry there are three Pt atoms (two on opposing sides of the trough, and the extra Pt atom in the trough) forming a mini monatomic wire spanning the trough, with three inequivalent adsorption sites located on it.\\
\indent The adsorption site with the highest binding energy of all
adsorption sites studied is the B$1$-site. In this case the CO molecule
binds to two Pt atoms through a bridge configuration. The distance
between the two Pt atoms is $2.682$ \AA, slightly longer than the
optimum inter-platinum-distance for a CO molecule adsorbed in a bridge configuration on a monatomic freestanding Pt wire. The B$2$- and B$3$-sites have an ontop configuration, and are $0.43$ eV and $0.25$ eV lower in adsorption energy. This shows a nice qualitative agreement between this 3-atom monatomic Pt wire present in the NW$2$-system and a freestanding monatomic Pt wire.\\
\indent For all adsorption sites studied on the NW$2$-surface, the CO bond lengths are only stretched slightly, $1.85-3.35$\%. Even more, comparing the values of \rco and \rcpt with those found by Sclauzero \textit{et al.} shows perfect agreement for the CO bond length, and just fractionally larger lengths for the C-Pt distances in our embedded 3-atom Pt wire. This slightly larger
length is a simple consequence of the Pt atoms being embedded in the
NW$2$-surface. The B$3$-site is the only site where the CO molecule is
perfectly perpendicular to the surface, this is due to the symmetry of
the chemical environment. In contrast, the B$2$ adsorption site has a
lower symmetry and the molecule bends along the asymmetry direction
toward the 3-atom monatomic Pt wire.\\
\indent Contrary to CO adsorbed on the NW$1$-surface, CO adsorbed on
the NW$2$-surface seems much less destructive. This is because the NW dimers
are anchored at their position by the presence of the
extra Pt atom. Only CO adsorbed at the B$1$- or B$3$-site modifies the
NW significantly. In both cases, the CO molecule has a bond to the extra
Pt atom, weakening the bonds with the NW dimers. Furthermore, just as
for the NW$1$-surface, the CO molecules are positioned between the NW
dimers, pushing them outward, away from the molecule. The combination of
these two effects is large enough to break the bonds between the NW dimers
and the extra Pt atom. Due to the periodic boundary conditions and the
size of our unit cell, the two NW dimers recombine to form a
tetramer-chain between the two copies of the CO molecule. In contrast to
the NW$1$-surface, the free NW dimers will not be able to block the
next equivalent adsorption site due to the presence of a NW dimer
anchored in place by its accompanying extra Pt atom.
\section{Discussion}\label{sc:discusion}
The results from the previous section show no clear-cut image for the
CO adsorption system based solely on calculated energies, which might
already be suspected from the history of CO adsorption on pure Pt
surfaces. However, direct comparison with experiment is possible by means of
calculated STM images. This method has
already been shown very successful at identifying correct adsorption
sites for CO, even when the calculated energies fail to do
so.\cite{Bocquet:ss96,Pedersen:cpl99} At the moment of writing only
very little experimental work is available, and the actual underlying NW
type is not always entirely clear. The earliest experimental work on
this system, by \"Oncel \textit{et al.}\cite{Oncel:ss06}, presents the adsorption of CO molecules on NWs separated $2.4$ \AA, which
based on our previous calculations (Ref.~\onlinecite{Vanpoucke:prb09NW}) leads to the assumption that these wires can be considered solitary NWs. In these experiments only one adsorption site was observed. It appears as a
protrusion at negative bias, and as a depression at
positive bias. Later work, by Kockmann \textit{et al.}\cite{Kockmann:prb08},
studied the adsorption of CO molecules on arrays of NWs spaced only $1.6$ \AA.
Based upon the results presented in Ref.~\onlinecite{Vanpoucke:prb09NW}, we assume these wires to have a NW$2$-geometry. Unlike \"Oncel \textit{et al.}, Kockmann and collaborators observed two adsorption sites, both
different from the one observed by \"Oncel \textit{et al.}\\
\indent Based on this we will compare our results for solitary wires with the
experiments of \"Oncel \textit{et al.}, and those for array-NWs with the
experiments of Kockmann \textit{et al.}

\subsection{CO adsorption on Pt induced nanowires}\label{ssc:dis_NW}
\begin{figure}[!tb]
\begin{center}
 \includegraphics[width=8cm,keepaspectratio]{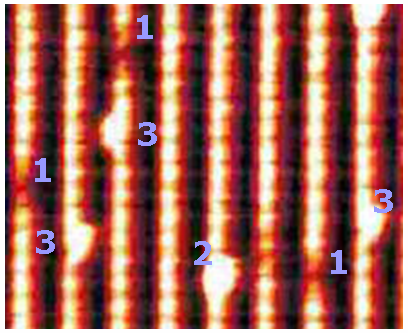}\\
\end{center}
  \caption{(Color online) Zoom out of a section of figure 1 in Ref.~\onlinecite{Kockmann:prb08}, an empty state STM image ($+$1.8 V; 0.5 nA) of CO-decorated NWs, recorded at 77 K. We have indicated three CO adsorption sites. Site 1 shows a large depression in the wire. Sites 2 and 3 show
  large protrusions. In case of site 2 nicely centered on the NW,
  while for site 3 the protrusion is centered either left or right of
  the NW.}\label{fig:3_Kockman_COsites}
\end{figure}

\begin{figure}[!tb]
\begin{center}
\includegraphics[width=8cm,keepaspectratio]{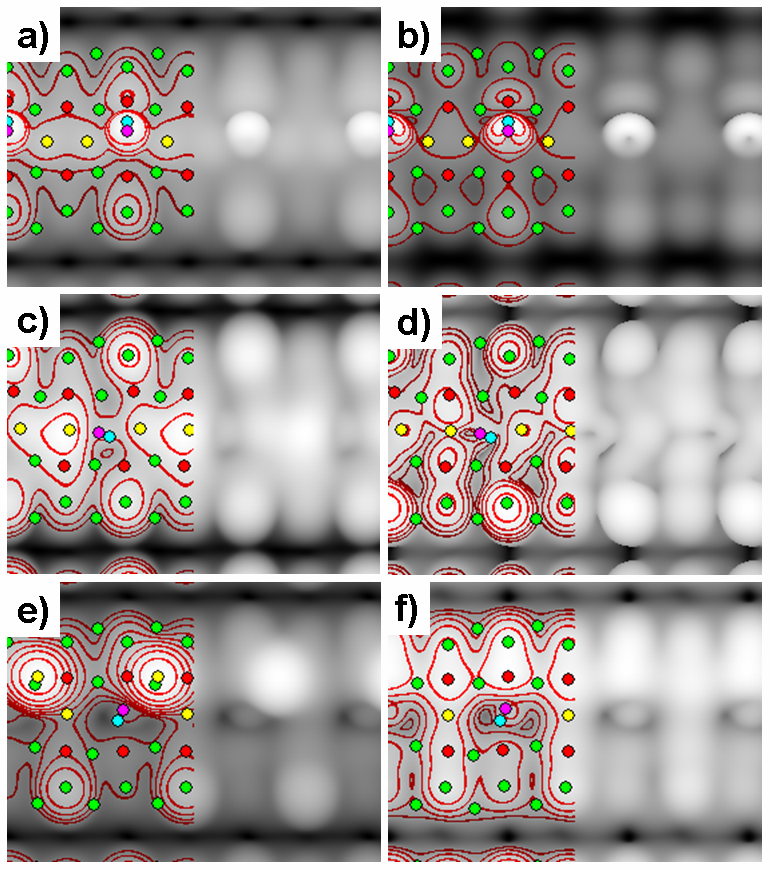}\\
\end{center}
  \caption{(Color online) Calculated filled and empty state STM images
  for the A$2$ (filled (a) and empty (b)), A$3$a (filled (c) and empty (d)), A$3$b (filled (e)), and A$3$b$1$ (filled (f)) adsorption geometries. The simulated biases are for (a), (e), and (f) $-1.50$ V,
  (b) $+1.50$ V, (c) $-0.70$ V, (d) $+0.30$ V. Contours are added
  to guide the eye, they are separated $0.3$ \AA. Colored discs indicate
  atomic positions: green and red represent Ge and Pt atoms in the top two
  layers of the surface, yellow represent the Ge atoms forming the NW,
  and cyan and fuchsia represent the C and O atoms respectively.
  }\label{fig:4_STM_A2_A3a}
\end{figure}

\begin{figure}[!tb]
\begin{center}
\includegraphics[width=8cm,keepaspectratio]{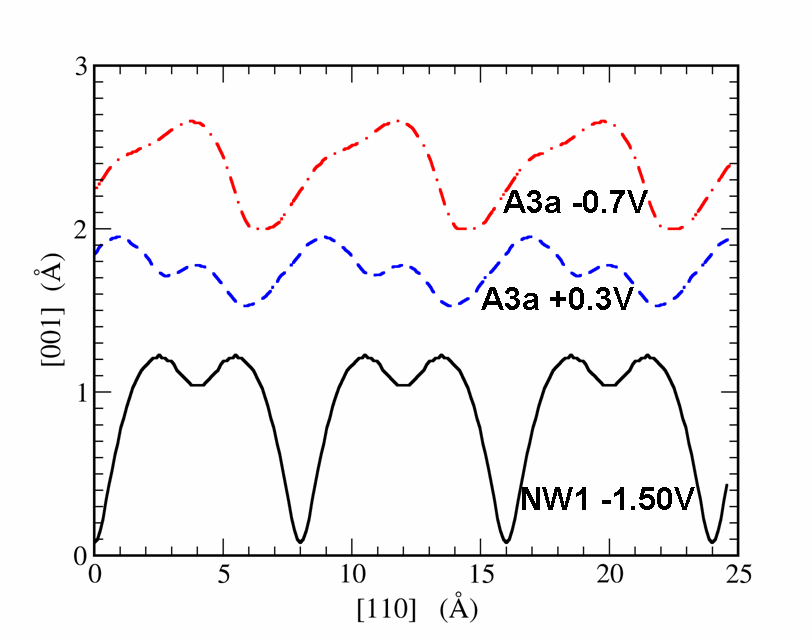}\\
\end{center}
  \caption{(Color online) Linescan images along the NW for the clean NW
  at a simulated bias of $-1.50$ V (solid black curve), the CO molecule
  adsorbed at the A$3$a-site at a simulated bias of $-0.70$ V (dash-dotted red curve) and $+0.30$ V (dashed blue curve). For each system $z=2.5$ \AA\ is used to generate the calculated STM images from which the line scans are taken. The lines are shifted along the [001]-axis.}\label{fig:5_LS_A2_A3a}
\end{figure}

\begin{figure}[!tb]
\begin{center}
\includegraphics[width=8cm,keepaspectratio]{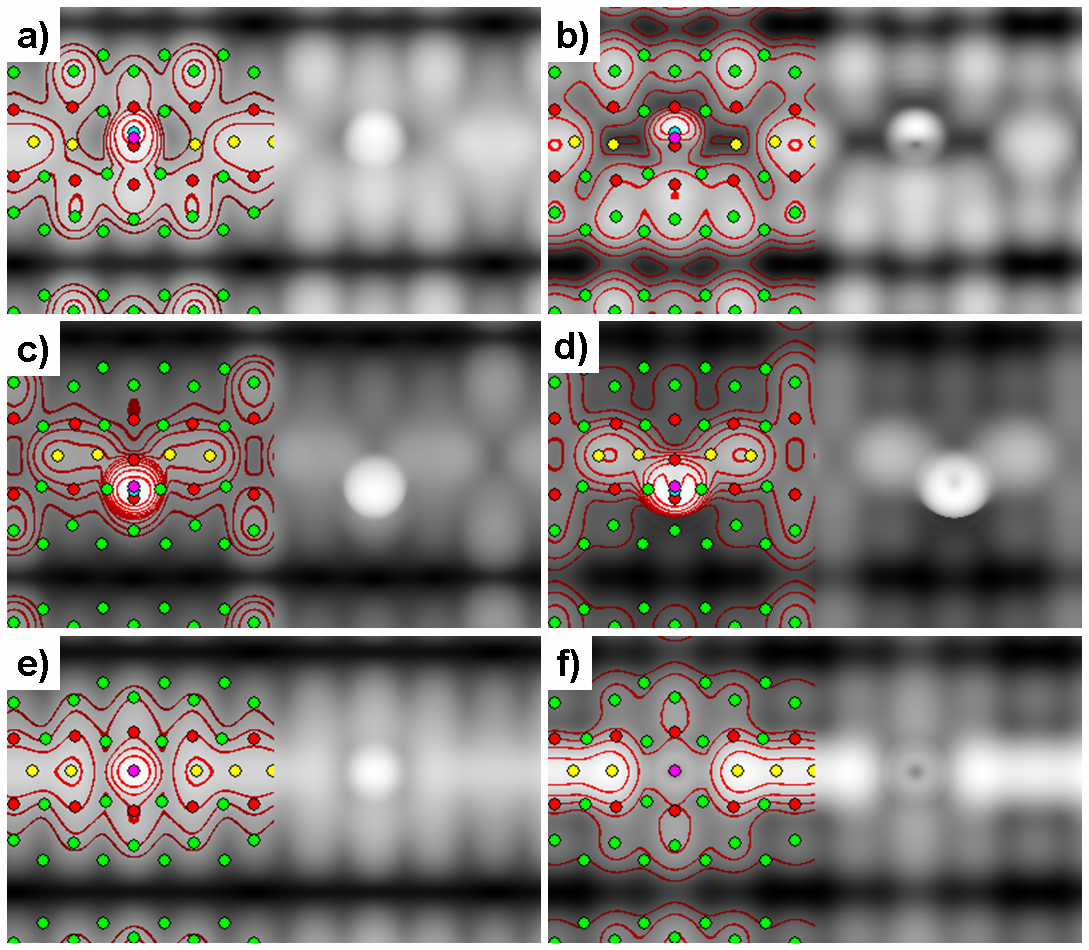}\\
\end{center}
  \caption{(Color online) Calculated STM images of CO molecules adsorbed at the
  B$1$- (a and b), B$2$- (c and d) and B$3$-sites (e and f). The distance
  above the highest atom was chosen $z=2.0$ \AA. Filled state images (a), (c)
  and (e) are at a simulated bias of $-1.80$ V, while for the empty state images (d) and (f) a simulated bias of $+1.80$ V was used. For the empty state image (b) a simulated bias of $+1.50$ V was used. Contours are added to guide the eye. All contours are separated $0.2$ \AA\ in the $z$ direction.
  Colored discs indicate the atom
  positions in the top layers. Ge and Pt atoms in the two top layers
  are shown in green and red respectively. Yellow discs are used to
  indicate the Ge NW atoms, while cyan and fuchsia is used for the C
  and O atoms respectively.}\label{fig:6_STM_B1_B2_B3t}
\end{figure}

As reference images for the experiment we will use figure $2$a and $2$b of
Ref.~\onlinecite{Oncel:ss06} for the solitary NWs and figure $1$ of Ref.~\onlinecite{Kockmann:prb08} for the NW arrays. Close examination of figure $1$ in Ref.~\onlinecite{Kockmann:prb08} shows three distinctly different adsorption sites, indicated in Fig.~\ref{fig:3_Kockman_COsites}. Kockmann and collaborators on the contrary only identify two, considering sites $2$ and $3$ the same adsorption site.
It is well known from literature that the electric field of the STM tip
can influence the position and orientation of molecules adsorbed on a
surface.\cite{Ramos:jp1991} This can result in an extra broadening of
the CO image along the scan lines.\cite{fn:fig3scanlines} This, however, can not cause the CO molecules at the adsorption site
$2$ in Fig.~\ref{fig:3_Kockman_COsites}, to appear as site $3$, since
the latter is observed at both sides of the NW [\textit{cf.}\ Fig.~\ref{fig:3_Kockman_COsites}].\\
\indent In addition to the formation and adsorption energies, STM images
are calculated for all adsorption NW$1$-and NW$2$-geometries. All NW$1$ CO-adsorption
geometries show protrusions for both positive and negative simulated
bias, with the exception of the A$3$a adsorption site. For the latter, a
protrusion is visible in the filled state image [\textit{cf.}\ Fig.~\ref{fig:4_STM_A2_A3a}c], while a depression is clearly present in the empty state image [\textit{cf.}\ Fig.~\ref{fig:4_STM_A2_A3a}c].
The comparable A$3$b-structure on the other hand shows a clear protrusion
on the adjacent quasi-dimer row (QDR) for all biases [\textit{cf.}\ Fig.~\ref{fig:4_STM_A2_A3a}e]. However, this protrusion
is not caused by the CO-molecule present, but by the Ge NW atom
ejected from the trough instead [\textit{cf.}\ Fig.~\ref{fig:2A3akat3geom}b].
Removal of this Ge atom removes the protrusion and only a brightened Pt-Ge dimer, bound to the O atom, remains [\textit{cf.}\ Fig.~\ref{fig:4_STM_A2_A3a}f]. The CO molecule itself remains invisible.\\
\indent Of all adsorbed CO molecules only those on the A$3$ sites and on the Ge NW, are located ``on'' the NW. However, these are not the only CO
molecules resulting in a CO-image ``on'' the wire. Due to the large tilt
angle of the CO molecule at the A$2$-site the resulting image gives the
impression of a CO molecule sitting, just slightly asymmetric, on top
of the NW, as can be seen in Fig.~\ref{fig:4_STM_A2_A3a}a and ~\ref{fig:4_STM_A2_A3a}b. For large negative bias the image is round, while becoming more and more bean-shaped for smaller negative biases and all positive biases.\\
\indent CO molecules bound to Ge NW atoms show images which look bean-shaped for the A$5$ adsorption site, and two-lobbed donut-shaped for the A$6$ adsorption site.\\
\indent Based on the adsorption energies found in
Sec.~\ref{sc:results}, the adsorption of CO on the Ge NW dimers can
be excluded. Only the Pt adsorption sites near the NWs remain. For
the solitary NWs the A$3$b-site can be excluded because for both
positive and negative bias a depression is found, which is not reported in experiments.\\
\indent The sites A$1$, A$2$, and A$4$ show a
comparable behavior. Both in the filled and empty state pictures a
large protrusion is clearly visible, slightly asymmetric to the NW
position. Figures~\ref{fig:4_STM_A2_A3a}a and~\ref{fig:4_STM_A2_A3a}b show an A$2$ adsorbed CO molecule as example.\\
\indent The A$3$a-site despite its low adsorption energy shows
something interesting. The filled state picture shows a pear shaped image for the NW dimer [\textit{cf.}\ Fig.~\ref{fig:4_STM_A2_A3a}c], just as was observed by \"Oncel \textit{et al.}, and the empty state picture shows no NW dimer image [\textit{cf.}\ Fig.~\ref{fig:4_STM_A2_A3a}d]. Comparison to figures $2$a
and b in Ref.~\onlinecite{Oncel:ss06} shows good agreement.
Linescans along the NW for the A$3$a
adsorption site, in comparison to a pristine NW are shown in Fig.~\ref{fig:5_LS_A2_A3a}. Due to periodic
boundary conditions and the fact that our unitcell only contains $1$ NW
dimer, this image can only be used to compare to the region $\sim0.5$
-- $\sim1.5$ nm in figure $3$ of Ref.~\onlinecite{Oncel:ss06}. Note that the linescans in figure $3$ of Ref.~\onlinecite{Oncel:ss06} and Fig.~\ref{fig:5_LS_A2_A3a} of this work are mirror images of each other (mirrored around the center of a NW dimer), indicating that the CO molecule was bound (the O-Ge bond) to the `left' side of the NW dimer in experiment, while it is bound to the right side in our calculations [\textit{cf.}\ Fig.~\ref{fig:4_STM_A2_A3a}c]. For the linescan of the filled state picture, the asymmetric shape and the width of the protrusion match very well.
There is also a good agreement between the linescans of the empty state
pictures. Note that in both cases the two small peaks have a different height,
with the highest peak at the lower side of protrusion in the filled state
picture. Furthermore, comparison to the linescan of a pristine NW shows
the maximum of the filled state protrusion to be located near the
center of the dimer, giving the impression that the CO molecule is
located on top of the NW dimer in a bridge configuration (compare to
figure $6.1$b in Ref.~\onlinecite{Oncel:thesis}). Also the
location of the larger of the two small protrusions, in the empty state
picture linescan, at the minimum between two NW dimers is in excellent
agreement with the experimental observations (\textit{cf.}\ figure $6.2$b in Ref.~\onlinecite{Oncel:thesis}). This shows that a CO molecule in between
two NW dimers, bound to one NW dimer through the O atom, can look like a molecule bound on top of a NW dimer. The low binding energy found here remains
problematic. However, the fact that taking into account the stretching
of the molecule returns an energy comparable and better than most of the other adsorption structures could indicate the energy barrier for desorption to play an important role. This also indicates that for RT experiments where a much lower CO density is present, reducing the contributions of direct and indirect interaction between CO molecules, this might not be as problematic. The good agreement of this structure with the experiment seems to support this idea.\\
\indent Although \"Oncel \textit{et al.} do not report observing any other
adsorption sites, they do report the CO molecules to perform a $1$D random walk along the NW. Since the A$3$a adsorption site does not easily allow
for a CO molecule to just jump from one site to the next, some
intermediate stable adsorption sites should be present to accommodate
this mobility. Looking at the geometry of the relaxed structures a path
can be imagined going from A$3$a to A$2$ [\textit{cf.}\ Fig.~\ref{fig:1NWadsorbsites}a], by breaking the bond between
the C atom and the Pt atom at the bottom of the trough and breaking the
O-Ge bond. Rotation from the A$2$ to the A$4$ configuration and onto
the A$1$ adsorption site, followed by the same path in reverse to the
next A$3$a site. The binding energies of these three adsorption sites
(A$1$, A$2$ and A$4$) differs only little making it an energetically
possible path at RT.\cite{fn:compcost} These sites are also present on the NW$2$-surface, where the adsorption site B$2$ can be considered an alternative for the A$2$-site. The calculated STM images of the B$2$-site also show the same asymmetric protrusion in the filled and empty state pictures, while the
adsorption energy (shown in Table~\ref{table:3energeom_NW2}) lies in
the range of the three A-sites. Site $3$ in Fig.~\ref{fig:3_Kockman_COsites}
clearly shows such an asymmetric adsorption site.
Although the resolution in Fig.~\ref{fig:3_Kockman_COsites} is not sufficient to distinguish between the four adsorption sites mentioned above, it is sufficient to indicate their existence.\\
\ \\
\indent For the NW$2$-surface, the calculated STM images for the adsorption sites show quite a complex picture. The most simple behavior is observed for the B$4$-site. Both the filled and empty state
pictures show, a round, slightly asymmetric CO image sticking out
far above the surface and the NW. Conversely, a CO molecule at the B$1$-site shows a sharp round image which becomes smaller (even invisible) for biases
close to the Fermi level [\textit{cf.}\ Fig.~\ref{fig:6_STM_B1_B2_B3t}a and~\ref{fig:6_STM_B1_B2_B3t}b]. CO adsorbed at the B$3$ site, on the other
hand, shows a nice round image for negative bias (\textit{cf.}\ filled state picture in Fig.~\ref{fig:6_STM_B1_B2_B3t}e), which becomes a two lobbed image for small positive bias but becomes invisible for large positive bias (\textit{cf.}\ empty state picture in Fig.~\ref{fig:6_STM_B1_B2_B3t}f).\\
\indent In each of the above cases the CO image appears nicely centered on the NW, which can be understood from the underlying geometry. In case of a CO molecule adsorbed at the B$2$-site ,the tilting of the molecule over the trough causes the CO image to appear only slightly shifted away from the
center of the NW, giving the impression the CO molecule might be
located on the wire itself [\textit{cf.}\ Fig.~\ref{fig:6_STM_B1_B2_B3t}c and~\ref{fig:6_STM_B1_B2_B3t}d]. Both the filled and empty state pictures show nice elliptical CO images for large biases, while close to the Fermi level the elliptical image becomes two-lobbed. As with all the previous cases where a donut or two-lobbed CO image was observed, this are the $\pi$-orbitals of the molecule that are being observed.\cite{Bocquet:ss96}\\
\indent The only adsorption site left to discuss, is the B$5$-site. Here the
CO molecule bridges the entire trough, so a serious
modification of the calculated STM pictures might be expected. Amazingly, the
calculated STM images show \emph{nothing}. Both for filled and empty state
pictures a normal NW image is observed, and not even the slightest
indication of the CO molecules presence can be observed, making the CO
molecule effectively \emph{invisible}.

\begin{figure}[!tb]
\begin{center}
\includegraphics[width=8cm,keepaspectratio]{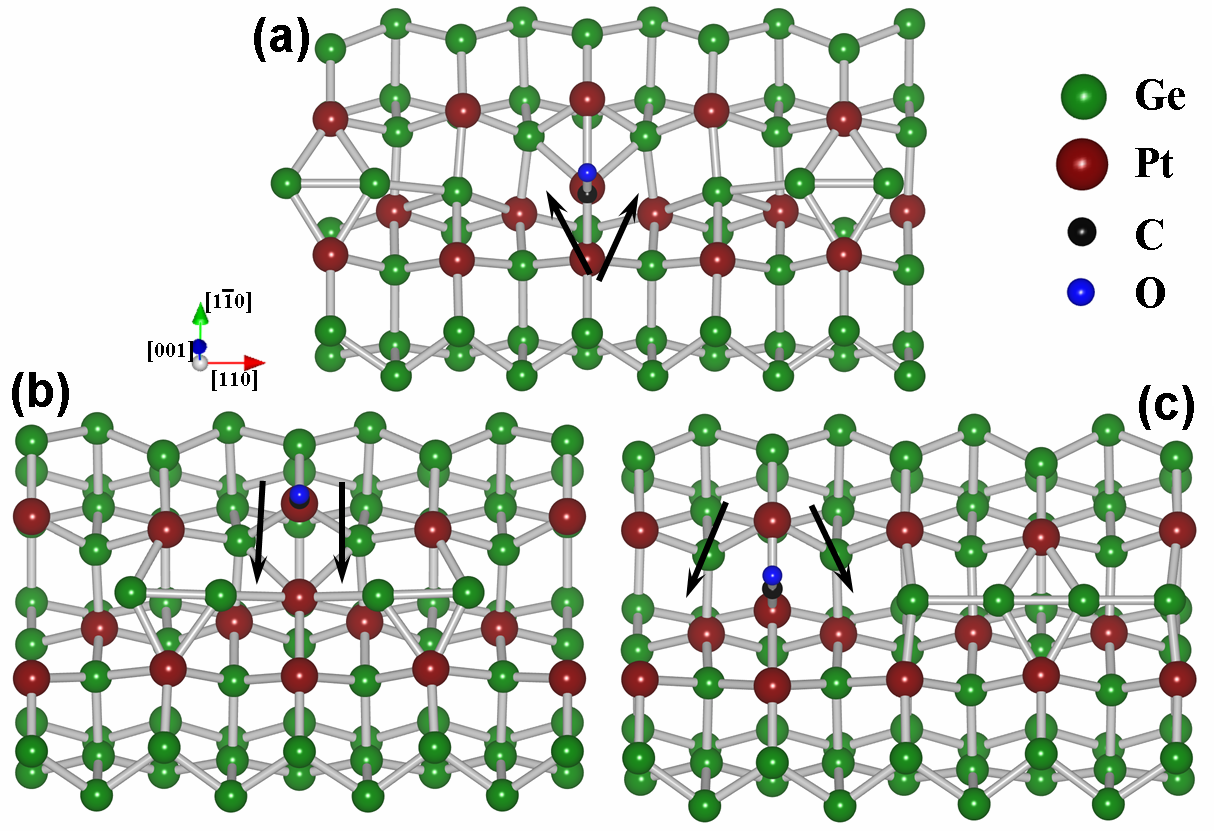}\\
\end{center}
  \caption{(Color online) Ball-and-stick representation of the relaxed
  geometries of CO molecules adsorbed at sites B$1$ (a), B$2$ (b) and
  B$3$ (c). Arrows indicate broken bonds between the Ge NW atoms and
  the extra Pt atom for the B$1$ (a) and B$3$ (c) adsorption sites. For
  the B2 adsorption site (b) these bonds, indicated with the arrows,
  are not broken.}\label{fig:7_geom_B1_B2_B3t}
\end{figure}
\indent Figure~\ref{fig:6_STM_B1_B2_B3t} shows the calculated STM
images for B$1$, B$2$, and B$3$ adsorbed CO molecules. When
comparing these images with experimental STM images, there are a few
things we need to keep in mind. i) The way the STM images were
calculated. By using a point source as tip, almost infinitely sharp
images are obtained, while in reality tip-size and geometry will
influence the obtained STM image. This will mainly manifest itself in a
broadening of the observed features. ii) No dynamic tip-substrate
interactions are included in the calculated STM images. Although for a
clean surface the effect of the tip on the surface geometry is almost
negligible, this is not the case for a molecule bound to the surface.\cite{Ramos:jp1991} At low coverage, molecules retain a large freedom
to move, even if their anchor point remains fixed, resulting in a
blurring of their observed STM image. iii) The position of the
molecular orbitals, especially the states above the Fermi level are not
that well described in DFT (this is the well known band gap problem). This means that it is not always possible
to use the same simulated bias as the experimental one. Points i and ii
explain why the CO images in Fig.~\ref{fig:3_Kockman_COsites} show up
to be sometimes two dimers long with a width larger than that of the NW.\\
\indent With this in mind, the three adsorption sites shown in
Fig.~\ref{fig:3_Kockman_COsites} can be identified by comparing the calculated
empty and filled state pictures to experimental STM pictures. Site $1$ shows
a depression in the empty state picture, centered between two NW dimers.
The filled state pictures on the other hand show a small
protrusion.\cite{Kockmann:prb08} Figures~\ref{fig:6_STM_B1_B2_B3t}e and~\ref{fig:6_STM_B1_B2_B3t}f show the same behavior for the B$3$ adsorbed CO molecule. Its geometry, shown in Fig.~\ref{fig:7_geom_B1_B2_B3t}c, shows the CO molecule bound in an on-top configuration to the extra Pt atom. This
bond weakens the bonds between the extra Pt atom and the NW dimers,
allowing for the presence of the CO molecule to break them entirely,
pushing the two NW dimers away from the CO molecule. The limited size
of the surface cell and the periodic boundary conditions results in the
formation of a Ge tetramer, and a limitation of the length of the depression to roughly $5$--$6$ \AA. In experiment however the NW
dimers could be pushed even further apart resulting in a large gap
around the CO molecule, which explains the experimentally observed
length of the depression to be roughly two dimer lengths. The lack of small
depressions on either side of the CO molecule in the experimental filled state
pictures can be understood as a consequence of the molecule-tip interactions mentioned earlier.\\
\indent The second site seen in Fig.~\ref{fig:3_Kockman_COsites}, shows
a protrusion in both filled and empty state pictures. Figures $3$a and
b in Ref.~\onlinecite{Kockmann:prb08} also show that the relative
height with regard to the NW is smaller in the empty state picture than
in the filled state picture. This turns out to be in agreement with the
images found for the B$1$ adsorbed CO molecule [\textit{cf.}
Fig.~\ref{fig:6_STM_B1_B2_B3t}a and~\ref{fig:6_STM_B1_B2_B3t}b]. At this adsorption site the CO molecule is bound in a bridge configuration to the extra Pt atom and a Pt atom of the surface dimer row [\textit{cf.} Fig.~\ref{fig:7_geom_B1_B2_B3t}a]. Again the bond with the extra Pt atom allows for the bonds between the NW dimers and the extra Pt atom to be broken, and the NW dimers to move away from the CO molecule resulting in a large gap around the CO molecule. Table~\ref{table:3energeom_NW2} shows this bridge configuration to have the highest adsorption energy, in agreement with the \textit{ab initio}
calculations of Sclauzero \textit{et al.}\cite{Tosatti:prb08}\\
\indent The third and last adsorption site indicated in
Fig.~\ref{fig:3_Kockman_COsites} shows a clearly  asymmetric protrusion in
the empty states picture. Unfortunately no experimental filled state pictures
have been published for this adsorption site, but based on all the other
adsorption sites observed in experiment we will assume that also in this case a protrusion is observed in the filled state image.
Figures~\ref{fig:6_STM_B1_B2_B3t}c and~\ref{fig:6_STM_B1_B2_B3t}d, show the filled and empty state pictures of a CO molecule adsorbed at site B$2$. This CO molecule is bound to a Pt atom in the surface dimer row, see
Fig.~\ref{fig:7_geom_B1_B2_B3t}b. Because it is tilted toward the NW, the
resulting image appears just slightly asymmetric of the NW, making it a
very good candidate for the adsorption site $3$ indicated in
Fig.~\ref{fig:3_Kockman_COsites}. In combination with the adsorption sites
A$1$, A$2$, and A$4$, this adsorption site gives a possible migration path for the mobility observed by \"Oncel \textit{et al.}\cite{Oncel:ss06}\\
\indent In their investigation of CO adsorption on the Pt induced NWs
Kockmann \textit{et al.} also observe a, what they call,
\emph{remarkably long-ranged repulsive interaction} between the CO
molecules. This repulsion, they found, has a range up to $3$--$4$nm (or $4$--$5$ NW dimers) along the NW direction. Due to its long range they concluded that this repulsive interaction can not just be a mere electrostatic repulsion. Furthermore, Kockmann \textit{et al.} note that the characteristic long-ranged repulsive interaction is independent of the adsorption sites involved. This long range interaction along the NW is sharp contrast with the fact that no significant interaction is observed between CO molecules on adjacent wires. This means the origin of the repulsive interaction needs to be linked to the NW itself. We have shown for the adsorption sites on the NW$1$- and the NW$2$-surface that the presence of CO molecules modifies the
nearby NW dimers in varying degrees. When the CO molecule is located
between NW dimers (\textit{e.g.} A$3$, B$1$, and B$3$) it seems to repel the
nearby NW dimers. For example, for the A$3$-site we find the resulting
modifications to extend up to two NW dimers in each direction. Two other examples are shown in Fig.~\ref{fig:7_geom_B1_B2_B3t}a and c. They show the interaction between two periodic copies of a CO molecule on the NW$2$-surface.
These copies are separated $2$ NW dimer apart and can press their
neighboring NW dimers toward one-another far enough such that they form
a tetramer. This effectively results in an indirect interaction between
the two CO molecules. The surface strain mediating the indirect CO interaction is directed purely along the NW itself. The NW dimers which are pressed away from their original position will in their turn press further neighboring NW
dimers from their equilibrium position and so on. These dislocated NW
dimers block the possible adsorption sites resulting in an
effective long range repulsive interaction.\\
\begin{figure}[!tb]
\begin{center}
\includegraphics[width=8cm,keepaspectratio]{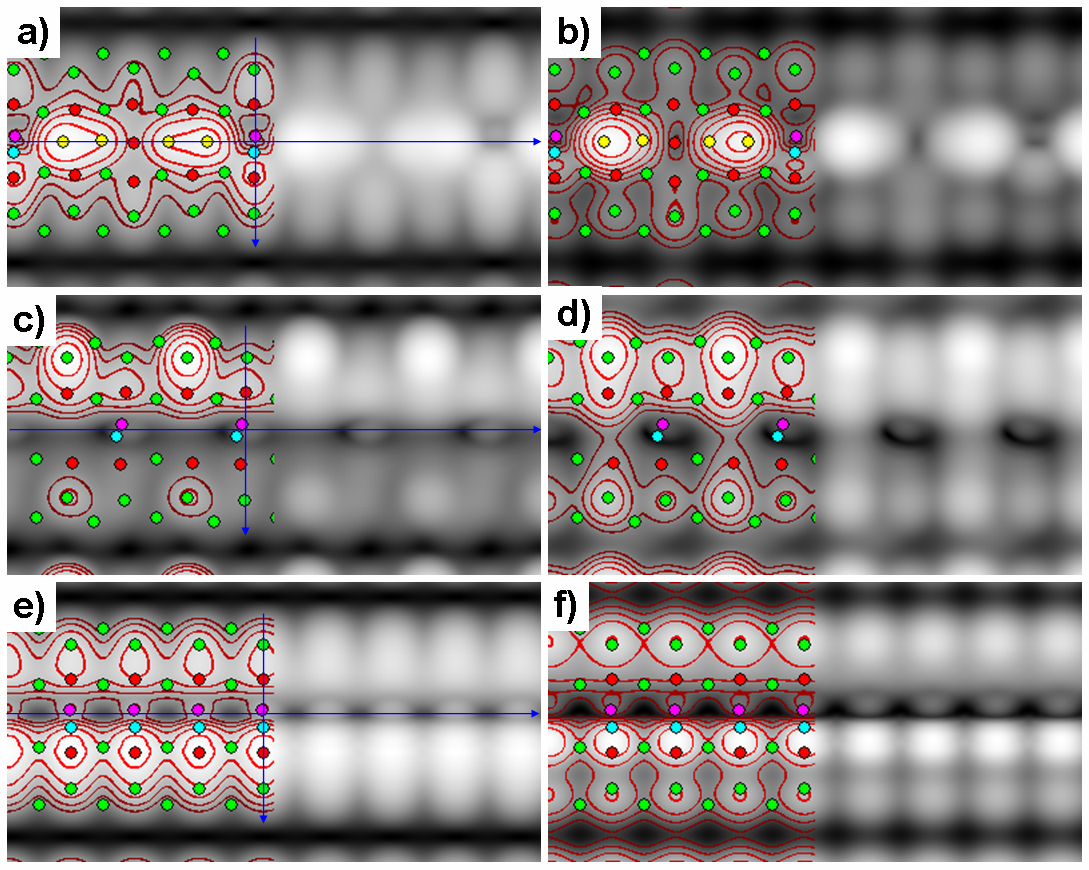}\\
\end{center}
  \caption{(Color online) Calculated STM images of CO molecules adsorbed in bridging
  configurations B$5$ (a and b), A$3$b$2$ (c and d) and A$7_{2\mathrm{CO}}$ (e and f). The distance above the highest atom was chosen $z=2.0$ \AA. Filled state pictures (a), (c) and (e) are at a simulated bias of $-1.50$ V, while for the empty state pictures (b), (d) and (f) a simulated bias of $+1.50$ V is used. Contours are added to guide the eye, and are separated $0.2$ \AA. Colored discs indicate the atom
  positions in the top layers. Ge and Pt atoms in the two top layers
  are shown in green and red respectively. Yellow discs are used to
  indicate the Ge NW atoms, while cyan and fuchsia is used for the C
  and O atoms respectively. Blue arrows show the position and direction
  of the linescans shown in Fig.~\ref{fig:10_LS_B5_A3b2_A7}}\label{fig:9_STM_B5_A3b2_A7}
\end{figure}
\indent As a final remark in this paragraph, we would like to point out
the B$5$ adsorption configuration. Its high adsorption energy (only
$140$ meV below that of the B$1$ configuration) makes it also a
reasonable adsorption configuration. Figures~\ref{fig:9_STM_B5_A3b2_A7}a and~\ref{fig:9_STM_B5_A3b2_A7}b show the calculated filled and empty state pictures. These show the CO molecule to be invisible, while the rest of the NW image remains unchanged. This is also the case for other simulated biases
($\pm0.30$ and $\pm0.70$ eV), leading to the conclusion that this
adsorption configuration could well be present in experiment, but
invisible for STM. Only high resolution linescans orthogonal to the NWs at the
position indicated in Fig.~\ref{fig:9_STM_B5_A3b2_A7}a could show the induced
asymmetry of the QDRs [\textit{cf.} Fig.~\ref{fig:10_LS_B5_A3b2_A7}], possibly combined with the observation that these asymmetric linescans are on average lower than those symmetric ones on locations where no CO molecule is present. Another option one could imagine to make the CO molecule appear, is by breaking its O-Pt bond with a small current surge through the STM tip. The CO molecule might then revert to the B$2$ adsorption configuration, which would be clearly visible in both filled and empty state pictures [\textit{cf.}\ Fig.~\ref{fig:6_STM_B1_B2_B3t}c and~\ref{fig:6_STM_B1_B2_B3t}d].\\
\indent This was for the case where the substrate is a NW array. For solitary
NWs, which we have already shown to be less stable under CO adsorption,
observation might be a little bit easier. Table~\ref{table:4_1DmolEl}
shows the A$7$ configuration (\textit{cf.}\ Fig.~\ref{fig:8_bridgesites}b, it
is the analog of the B$5$ configuration) to have a very high adsorption
energy. Here the steric repulsion between the CO molecule and
the NW dimers will create a hole in the NW centered around the CO
molecule. This defect should show up as a depression over a wide range
of biases. This wide depression would have a small protrusion in its
center and a width of at least $8$ \AA.\cite{fn:widthA7depri}
\begin{table*}[!t] \center{\textbf{Adsorption energies and geometrical
parameters for bridging CO molecules\newline on a Pt
modified Ge(001) surface.}}
\begin{ruledtabular}
\begin{tabular}{l|rrcrcrcr}
   & \multicolumn{1}{c}{\Ef} & \multicolumn{1}{c}{\Ead} &
   \multicolumn{1}{c}{coord.} & \multicolumn{1}{c}{\rco} &
\multicolumn{1}{c}{\aco} & \multicolumn{1}{c}{\rcpt} &
\multicolumn{1}{c}{\acpt} & \multicolumn{1}{c}{\ropt} \\
   & \multicolumn{1}{c}{(eV)} & \multicolumn{1}{c}{(eV)} & &
   \multicolumn{1}{c}{(\AA)}&
\multicolumn{1}{c}{($^{\mathrm{\circ}}$)} & \multicolumn{1}{c}{(\AA)} &
\multicolumn{1}{c}{($^{\mathrm{\circ}}$)} &
\multicolumn{1}{c}{(\AA)}\\
  \hline
  NW$1$ A$3$b& $-1.485$  & $-2.069$ & b & 1.214 & 47 & 2.012 & 101 & 2.249
  \\
             &         &        &   &       &    &(2.066)&     \\
  NW$1$ A$3$b1& $-1.163$ & $-2.276$ & b & 1.207 & 43 & 1.960 & 90 & 2.167
  \\
             &         &        &   &       &    &(2.106)&     \\
  NW$1$ A$3$b2& 0.390  & $-2.094$  & b & 1.204 & 44 & 1.964 & 87 & 2.190
  \\
             &         &        &   &       &    &(2.084)&     \\
  NW$2$ B$5$ & $-0.784$  & $-2.197$  & t & 1.170 & 20 & 1.883 & 28 & 2.433 \\
  NW$1$ A$7$ & $-1.589$  & $-2.361$  & t & 1.169 & 17 & 1.884 & 28 & 2.401 \\
  NW$1$ A$7_{2\mathrm{CO}}$ & $-3.741$  & $-2.214$  & t & 1.167 & 19 & 1.882 & 27 &  2.394 \\ 
  \end{tabular}
\end{ruledtabular}
\caption{Formation and adsorption energies for CO adsorbed
in configurations bridging the trough between two neighboring QDRs.
Adsorption sites are shown in Fig.~\ref{fig:8_bridgesites}a. The B5
and A3b values are duplicated from Table~\ref{table:1energeom_NW1}
for ease of comparison. $\mathbf{r}_{\mathrm{CO}}$, $\mathbf{r}_{\mathrm{C-surf}}$, and \ropt are the C-O, C-Pt and
O-Pt bond lengths, the value between brackets for \rcpt is the bond
length of the C atom to the Pt atom at the bottom of the trough. \aco
and \acpt are the bond angles with regard to the surface plane. In case
of bridge adsorption \acpt is the angle between the two C-surface
bonds.}\label{table:4_1DmolEl}
\end{table*}
\begin{figure}[!tb]
\begin{center}
\includegraphics[width=8cm,keepaspectratio]{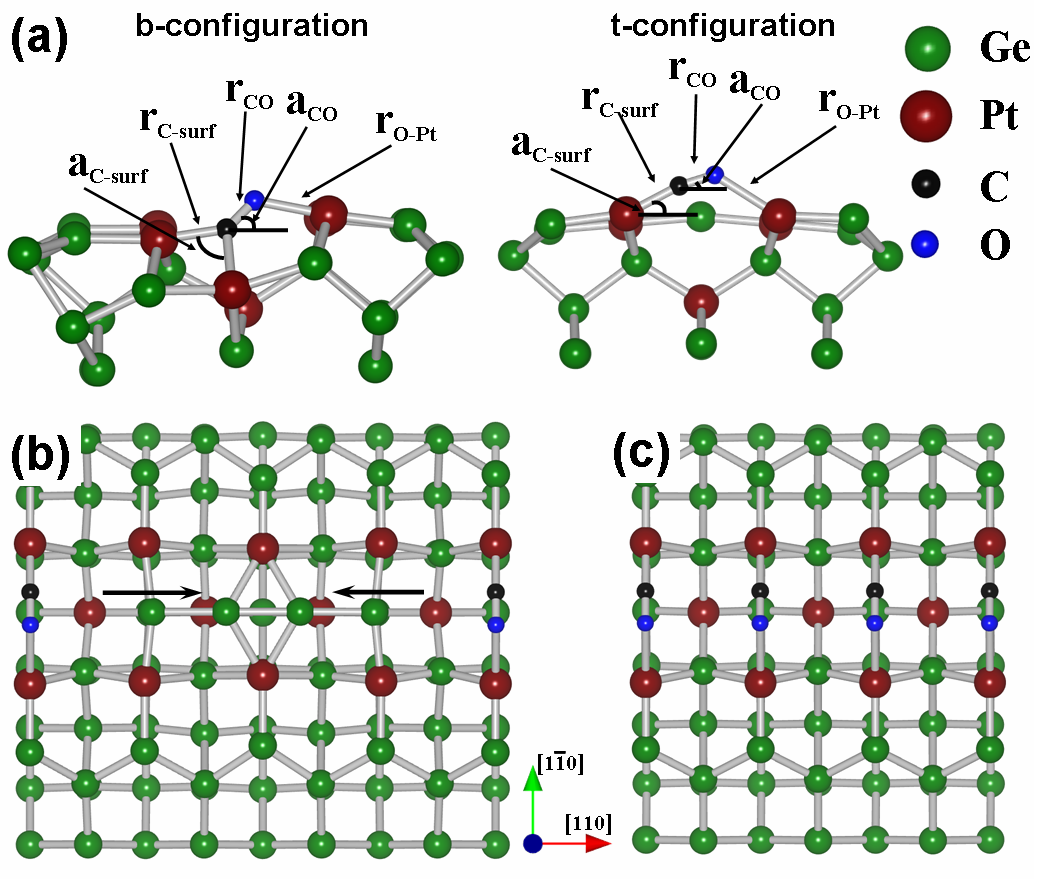}\\
\end{center}
\caption{(Color online) (a) Ball-and-stick representation of the two
bridging CO configurations. Geometrical parameters given in
Table~\ref{table:4_1DmolEl} are indicated. Ball-and-stick representation
of the A$7$ (b) and A$7_{2\mathrm{CO}}$ (c) structures.
Arrows in (b) show the drift direction of the NW
dimers.}\label{fig:8_bridgesites}
\end{figure}
\subsection{Molecular electronics on Pt modified Ge(001)?}
From these calculations and the experiments presented in literature, the
possible application of this system for $1$D molecular electronics
becomes very unlikely. The long-ranged interaction observed by Kockmann
\textit{et al.} and its explanation in light of the calculations performed in
this work, seems to be the main problem.\\
\indent Up to this point we mainly focussed on identifying the
experimentally observed structures. The B$5$ adsorption configuration,
and the comparable A$3$b configuration have not yet been discussed in light of experiments, while showing almost the best adsorption energies. The calculated STM images for these structures show something very peculiar: the total lack of an image for the CO molecule. In case
of the A$3$b adsorption site a large protrusion is still visible,
however this is an ejected Ge atom of the NW dimer [\textit{cf.}\
Fig.~\ref{fig:2A3akat3geom}b]. This ejection was due to the limited
unitcell size. The steric repulsion between two periodic copies of the CO molecule and the single Ge NW dimer prohibited the Ge dimer to be displaced sufficiently along the NW direction. To remove this `computational artefact' the ejected Ge atom is removed (A$3$b1), and in a second calculation
both Ge atoms forming the NW dimer are removed (A$3$b2). Also an
adsorption structure with B$5$ configuration on a NW$1$ surface is build
(A$7$), using a double NW$1$ surface cell.
\begin{figure}[!tb]
\begin{center}
\includegraphics[width=8cm,keepaspectratio]{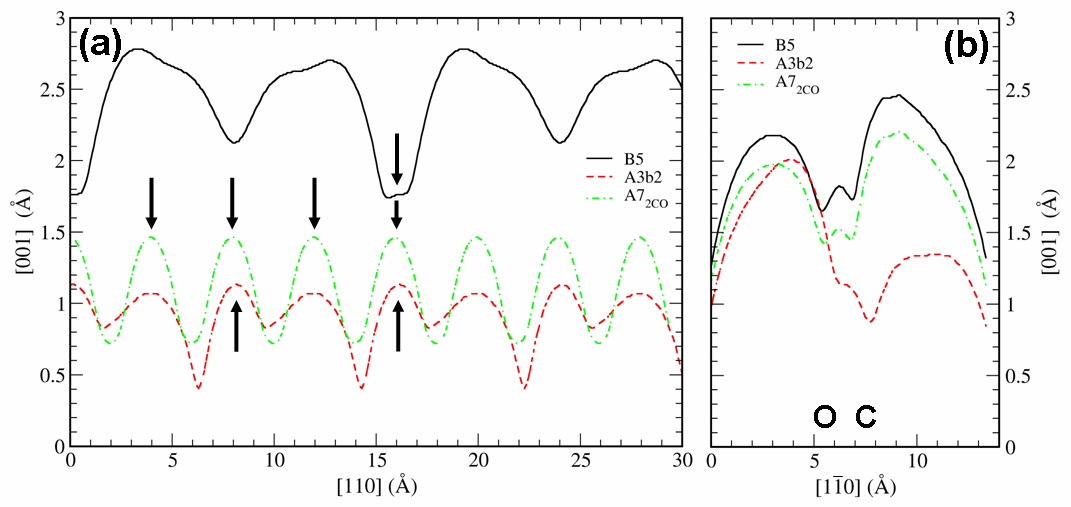}\\
\end{center}
  \caption{(Color online) Calculated linescan images along (a) and orthogonal (b) to the
  trough/NW for the B$5$ (solid black curve), the A$3$b$2$ (red dashed curve) and the A$7_{2\mathrm{CO}}$ (green dash-dot curve) configuration. The linescans are taken along
  the arrows shown in Fig.~\ref{fig:9_STM_B5_A3b2_A7} (a), (c), and (e).
  The arrows in (a) indicate the positions of the CO molecules, while C
  and O show the positions of the C and O atom in (b).
  }\label{fig:10_LS_B5_A3b2_A7}
\end{figure}
Table~\ref{table:4_1DmolEl} shows the formation and adsorption energies
for these new structures in comparison to the B$5$- and A$3$b-structures. Removal of the Ge NW atoms clearly decreases the formation energy of the surface, which is due to the uncovering of the imbedded Pt atoms.\cite{Vanpoucke:prb09NW}
The adsorption energy of the CO molecule however remains roughly the
same. Also the geometrical parameters barely change. The bond lengths
decrease slightly, due to the reduced steric repulsion between
the CO molecules and the Ge dimer atoms, moving the CO molecule closer to the surface.\\
\indent In case of the A$7$ adsorption configuration, the lack of
anchor point for the NW dimers causes them to drift away from the CO
molecule due to the steric repulsion. Which, because of the limited
unit cell size, results in the formation of a Ge tetramer as is shown in
Fig.~\ref{fig:8_bridgesites}b. The geometric parameters for the CO molecules
in the B$5$ and A$7$ adsorption configurations are almost identical. However, the distance between a CO molecule and the nearest Ge atom of a NW dimer is approximately $4.3$ \AA\ in case of the A$7$ configuration, while it is only $3.3$ \AA\ in case of the B$5$ configuration. This indicates the improvement in adsorption energy, going from the B$5$ to the A$7$ configuration, can be attributed to the reduction of the steric repulsion between the CO molecule and the Ge NW dimers.\\
\indent Because uncovering the imbedded Pt atoms has a negative influence on the formation energy, a model is build with the entire NW removed and replaced by a maximum coverage of CO molecules in an A$7$ configuration (A$7_{2\mathrm{CO}}$).
The A$7_{2\mathrm{CO}}$ structure contains two CO molecules per $4\times2$
surface cell [\textit{cf.}\ Fig.~\ref{fig:8_bridgesites}c], making this a
four times higher coverage than the A$7$ case. A large increase in
the formation energy is found, while the steric repulsion only
has a minute influence on the adsorption energy of the CO molecules.
Furthermore, the geometrical parameters remain almost unchanged making
this, from the geometrical point of view, a very good candidate for
$1$D molecular electronics. However, one small problem remains: these
CO molecules in either b or t bridging configuration are invisible in
STM. Figure~\ref{fig:9_STM_B5_A3b2_A7} shows both filled and
empty state pictures of the B$5$, A$2$b$2$, and A$7_{2\mathrm{CO}}$
adsorption geometries. For the B$5$ adsorption configuration the
linescan (black solid line shown in Fig.~\ref{fig:10_LS_B5_A3b2_A7}) is almost
unmodified compared to the linescan of the pristine NW. The only modification
is found in the line scan orthogonal to the trough/NW. In case of a t
bridging configuration, the dimer of the QDR which is bound to the C
atom is higher than the dimer bound to the O atom, while the opposite
is true for the b bridging configuration.\\
\indent In conclusion, if one would succeed
in stripping away the Ge NW without damaging the underlying substrate,
a high coverage of CO molecules in bridging configuration could be
recognized by this asymmetry in the QDR images. The formation and
adsorption energies are highly favorable, making this also
energetically an interesting template for $1$D molecular electronics.\\

\section{Conclusions}\label{sc:conclusion}
In this paper we have studied the adsorption of CO molecules on Pt
induced NWs on Ge(001) using \textit{ab initio} DFT calculations. We
show CO has a strong preference for adsorption on the Pt atoms imbedded
in the Ge(001) surface. As a consequence CO molecules do not bind
directly on top of the Ge dimers forming the NWs, contrary to
the experimental assumptions. By direct comparison of calculated
STM images to experimental STM images we have successfully identified
the observed adsorption sites. We have shown that the Pt atoms lining
the troughs in which the Ge NWs are imbedded provide the necessary
adsorption sites to explain all experimentally observed CO adsorption
sites.\\
\indent CO molecules in ontop configurations next to the NW tilt toward
it, presenting STM images located on the NW. CO molecules bound in
between NW dimers, with the O atom also bound to a Ge NW dimer modify
the electronic structure of this Ge atom sufficiently to give the
appearance of a protrusion on this Ge dimer. This gives rise to the
short-bridge CO adsorption site observed by \"Oncel \textit{et al.} A
CO molecule bound in an ontop configuration on the extra Pt atom of
the NW$2$-surface, showing a protrusion at negative bias and a
depression at positive bias, is found to show good agreement with the
experimentally observed long-bridge site, seen by Kockmann \textit{et al.} The
short-bridge site observed by this group is identified as a CO molecule
in a bridge configuration in between NW dimers.\\
\indent A path for mobility along the wire is presented, showing the CO
molecule to move along the Pt atoms of the underlying QDRs.
The long-ranged interaction observed by Kockmann \textit{et al.} is
explained trough the dislocation of NW dimers, in the vicinity of the
CO molecule.These dislocated NW dimers in turn block the nearby CO-adsorption
sites.\\
\indent We also predict the presence of invisible bridging CO
molecules, and present methods for observing them experimentally.
Also the possibility of $1$D molecular electronics is touched. After removal of the Ge NW dimers, stable, invisible wires of parallel CO molecules, along the
Pt lined troughs, can be obtained. This configuration has
a large formation energy \Ef$=-3.74$ eV and an adsorption energy per
CO molecule \Ead$=-2.21$ eV at maximum CO coverage.
\section*{Acknowledgements}
This work is part of the research program of the ``Stichting voor
Fundamenteel Onderzoek der Materie" (FOM) and the use of supercomputer
facilities was sponsored by the "Stichting Nationale Computer
Faciliteiten" (NCF), both financially supported by the ``Nederlandse
Organisatie voor Wetenschappelijk Onderzoek" (NWO).



\begin{thebibliography}{37}
\bibitem[{fn:({\natexlab{b}})}]{fn:fig2surf}
\bibinfo{note}{Two surface unit cells are shown in
  Fig.~\ref{fig:2A3akat3geom}b.}
\bibitem[{fn:({\natexlab{d}})}]{fn:compcost}
\bibinfo{note}{The computational cost to calculate diffusion barriers between
  these intermediate steps is, due to the system size, too large.}
\bibitem[{fn:({\natexlab{e}})}]{fn:widthA7depri}
\bibinfo{note}{In our limited surface cell the NW dimers could not spread
  further before colliding with periodic copies, limiting the width of the
  depression to $\sim8$ \AA.}
\end{thebibliography}

\end{document}